# Single-mode and single-polarization photonics with anchored-membrane waveguides

**Jeff Chiles[1] and Sasan Fathpour[1,2,*]**

[1]*CREOL, The College of Optics and Photonics, University of Central Florida, Orlando, FL 32816*
[2]*Department of Electrical Engineering and Computer Science, University of Central Florida*
[*]*fathpour@creol.ucf.edu*

**Abstract:** An integrated photonic platform with "anchored-membrane" structures, the *T*-Guide, is proposed and numerically investigated. These compact air-clad structures have high index contrast and are much more stable than prior membrane-type structures. Their semi-infinite geometry enables single-mode and single-polarization (SMSP) operation over unprecedented bandwidths. Modal simulations quantify this behavior, showing that an SMSP window of 2.75 octaves (1.2 – 8.1 µm) is feasible for silicon *T*-Guides, spanning almost the entire transparency range of silicon. Dispersion engineering for *T*-Guides yields broad regions of anomalous group velocity dispersion, rendering them a promising platform for nonlinear applications, such as wideband frequency conversion.

## 1. Introduction

Modern integrated photonic platforms ought to achieve high optical confinement, low core and cladding material loss over a wide spectral range and polarization-state management. Wide transparency enables the same platform to be used for multiple applications requiring different wavelengths, taking full advantage of the compactness offered by an integrated approach. Polarization management is key to implementing sensitive on-chip measurements. One approach is to realize "single-mode single-polarization" (SMSP) waveguides, which deliberately restrict propagation to one mode of one polarization to improve the sensitivity in photonic devices [1]. The other polarization is leaked or radiated away by various means [2]. This eliminates the need for standalone polarizers and guarantees that no undesired polarization component is introduced to the system by fabrication imperfections.

Extensive work on the design and simulation of SMSP in the context of optical fibers has been conducted [3,4], with predicted SMSP windows of up to 2.62 octaves [5]. However, the experimentally achieved fibers have fallen far short of such figures with a maximum measured window of only 0.23 octaves [6]. This low value is partially due to the inherent complexity of perfectly realizing the intricate geometries of these structures.

In the context of integrated photonics, however, little work has been done on SMSP waveguides. To achieve compact, low-cost devices, operation on the centimeter scale is desirable, which is possible through microfabrication processing techniques. There are several integrated photonic applications that could benefit from SMSP operation. On-chip optical gyroscopes have the potential to provide reliable navigational data in a compact form-factor and at low cost. However, polarization fluctuations inherent to bipolarized waveguides (supporting modes of both polarizations) can impose limitations to their sensitivity [7]. With SMSP waveguides, this problem could be addressed readily. Another application for SMSP waveguides is for on-chip supercontinuum generation, which could improve polarization purity and beam quality by restricting propagation to a single mode. Spectra exceeding one or multiple octaves have been generated in standard waveguides with no polarization management [8,9]. Thus, the demand in SMSP bandwidth would be much greater than what has been achieved to date.

It is well-established that operation with only the transverse electric (TE) mode can be achieved in ridge waveguides, when they are shallowly etched [10,11]. Single transverse-mode operation can also occur in such shallow ridge waveguides with small enough widths. However, the trade-off to such a design is that only small index contrasts are achieved, resulting in large modes and poor confinement. Thus, this approach is not suitable for dense photonic integration or nonlinear applications requiring a small mode area.

Alternative approaches to single-polarization waveguides have been considered [12,13]. A type of pedestal silicon waveguide was proposed in [13], but the technique is limited to using waveguides close to the confinement limit. This, combined with the asymmetry of the structure, causes the TE-mode to cut off at relatively short wavelengths, giving a limited predicted SMSP spectrum from 1.43 to 1.84 μm (0.36 octaves) for a challengingly narrow pedestal width of 100 nm (simulated in-house with COMSOL™). Propagation losses greater than 19 dB/cm were measured for the fabricated samples in [13].

Clearly, waveguide platforms with much broader SMSP bandwidth and much lower propagation losses are required to be useful as in-line polarizers. In this work, we show extremely broad SMSP windows and ideal dispersion profiles can be achieved with a novel structure we call the *T*-Guide.

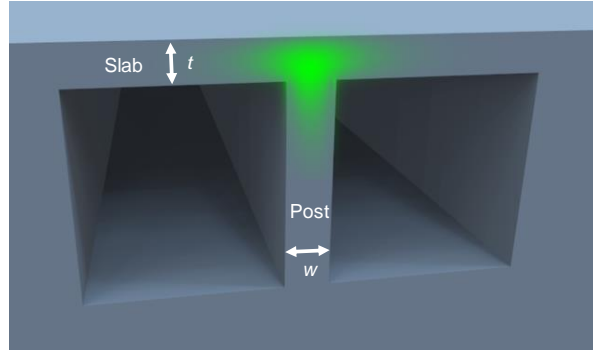

Fig. 1. Geometry of an anchored-membrane waveguide or *T*-Guide. The simulated intensity profile of a typical guided mode is overlaid at the junction between the post and the slab.

The proposed platform is schematically shown in Fig. 1. Resembling our prior demonstration of all-Si suspended-membrane waveguides [14], this structure can be readily fabricated; essentially, the processing consists of patterning a handle wafer with two trenches separated by a narrow post, followed by transfer of a thin layer of material (the "slab") on top, forming a guided mode at the junction between the post and slab. We have previously demonstrated low propagation losses of 1.75 dB/cm in silicon *T*-Guides at a mid-infrared (mid-IR) wavelength of 3.64 μm [15].

The unique geometry of the proposed structures results in exotic behavior such as ultra-broadband SMSP operation, as well as flexible dispersion engineering and precise control over mode shape and positioning. Also, since no cladding materials are needed, the transparency window is limited only by the core material chosen, making the structure ideal for tougher wavelengths such as the mid-IR [16]. This feature complements the *T*-Guides' ultra-broadband SMSP window and is a clear advantage over prior mid-IR Si platforms, such as silicon-on-sapphire [17] and silicon-on-nitride [18]. Additionally, a high index contrast can be retained in the waveguiding structure thanks to the air interfaces on all sides, allowing compact waveguide dimensions. In this work, the fundamental waveguiding behavior, SMSP operation, and dispersion engineering of silicon *T*-Guides are considered.

## 2. Properties of optical modes in *T*-Guides

*2.1 Waveguiding behavior*

To explain some of the unusual optical properties of *T*-Guides, it is helpful to first consider their fundamental waveguiding behavior. They bear some similarity to conventional ridge-type optical waveguides in that the optical mode is laterally confined by a defect in the slab thickness. But the guiding defect in this case is a post, which is semi-infinite in its height because it adjoins with the substrate. This produces markedly different optical properties from conventional ridge waveguides or truncated ridge-type waveguides, such as those simulated in [19]. In the same way that some ridge waveguides are able to "leak" and thus extinguish higher-order lateral modes [20], *T*-Guides are capable of doing so in both lateral and vertical directions, since the structure is semi-infinite in both directions. Unlike in ridge waveguides, however, widening the post of a *T*-Guide will not result in additional lateral modes, as they would simply be leaked to the substrate through the post. Similarly, increasing the slab thickness cannot result in additional vertical modes, since only one field lobe can occupy the junction where the local increase in effective index occurs. Numerical results presented in Section 2.2 confirm that no higher-order modes are present for any *T*-Guide design across the multiple octaves simulated.

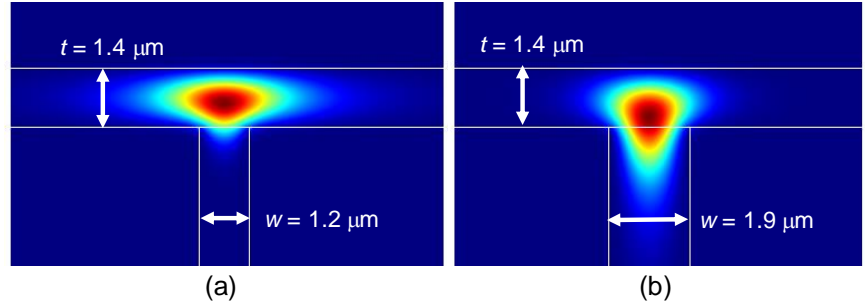

Fig. 2. Simulated intensity profile of the TE mode for different post widths (a) $w = 1.2$ μm; (b) $w = 1.9$ μm.

An intuitive understanding of the guiding behavior of *T*-Guides can be gained by considering it as a competition between the planar modes of the slab and those of the post. To illustrate this, we consider the quasi-transverse-electric (TE) guided modes (intensity distribution) in the structures of Fig. 2(a-b), which employ silicon as the core material, bounded by air on all sides. The confinement of the mode in the slab and post regions is determined by the effective index contrast between the guided mode and that of the respective planar mode in each region. For the slab region, this is a TE-like mode. For the post region, however, the electric field is oriented perpendicular to its long axis, so it experiences a TM-like planar mode index. The guided mode can thus be controlled in its size and shape by simply adjusting the relative dimensions of the slab and the post, which in turn changes the planar mode index contrast. For example, for a *T*-Guide operating at $\lambda = 3.6$ μm with a slab thickness, *t*, of 1.4 μm, we consider the effect of changing the post width, *w*, from 1.2 to 1.9 μm. The respective mode intensity profiles are provided in Figs. 2(a-b), while the effective index contrast between the slab and post modes is plotted versus the overlap of the optical mode in the post region in Fig. 3.

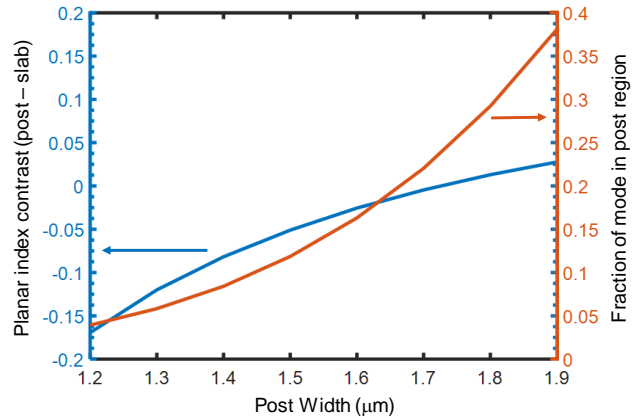

Fig. 3. Effect of increasing the post width on the position of the optical mode. A wider post increases the TM-like planar mode index in the post region, which increases the index contrast (left axis), resulting in a higher fraction of optical power in the post (right axis).

For the smaller *w*, the mode resides largely in the slab. But it can be seen that once the planar mode index of the post overtakes that of the slab (positive index contrast in Fig. 3), there is a rapid change in the mode distribution toward the post side. This provides a means of controlling the size and position of the optical mode in *T*-Guides. For instance, a narrow post width can be used to obtain wider modes necessary for efficient input and output coupling, or passive structures such as splitters and directional couplers. Reasonably compact waveguide bends can also be achieved with *T*-Guides; for example, an Si *T*-Guide with $w$ = 800 nm and $t$ = 500 nm experiences a critical bend radius of 190 μm, as determined through axially-symmetric eigenmode simulations. Since these waveguides are inherently single-mode, bends will never cause light to be lost to higher-order modes as would be the case in many conventional waveguide designs.

*2.2 SMSP properties*

In addition to their single-transverse-mode properties, *T*-Guides also exhibit single-polarization guidance over extremely broad bandwidths, and they can do so while operating in a well-confined state, far from cutoff. The broken symmetry of the structure in the vertical direction results in a rapid cutoff only for the vertically polarized, TM-like mode. To investigate this, a series of simulations were performed on various silicon *T*-Guide structures with air cladding on all sides. A simulated structure having a membrane width of 10 μm on either side, a trench depth of 4 μm, and a slab thickness $t$ = 500 nm was employed. COMSOL™ was used to examine any guided modes as the post width, $w$, was varied in several steps from 300 to 800 nm.

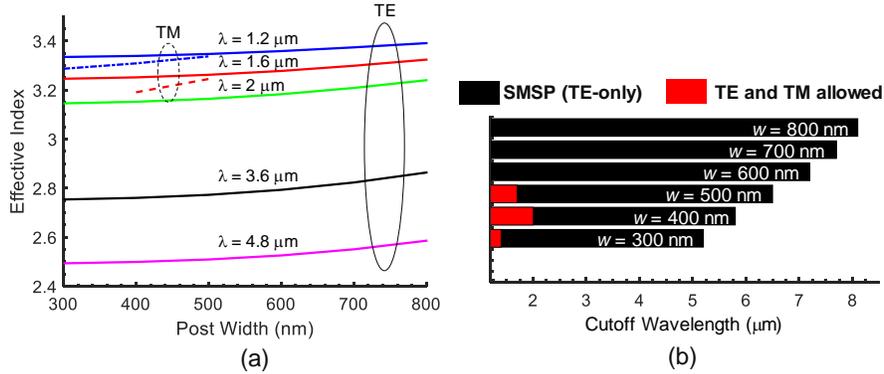

Fig.4. (a) Post width vs. effective index for various wavelengths. TM modes are plotted as dashed lines and TE as solid lines; (b) Guided mode transmission windows corresponding to several post widths, spanning from 1.2 μm up to the 10 dB/cm leakage loss cutoff for each case.

Scattering boundary conditions and a leakage loss threshold of 10 dB/cm were assumed to define the cutoff condition for modes. This value is reasonable, as it is generally chosen to correspond to the relevant length-scale of the platform (e.g., 1 dB/m in the case of [3]). The modal effective indices for different post widths are plotted in Fig. 4(a), along with the corresponding windows of operation defined by the cutoff condition in Fig. 4(b). No higher-order modes were present for any design at any of the simulated wavelengths of 1.2 to 8.5 μm, confirming the single-transverse-mode property of these structures. It can be seen that the TM modes are guided only over a very narrow bandwidth in the near-infrared, in the cases of $w$ = 300, 400 and 500 nm. Their transmission windows vary non-monotonically with post width due to different coupling to the slab. However, they completely vanish for $w \geq 600$ nm, and in the case of $w$ = 800 nm, pure TE-mode operation is enabled in the $\lambda$ = 1.2 - 8.1 μm range, which is nearly the entire transmission window of silicon. This represents an SMSP window of 2.75 octaves, a record span for both integrated and fiber-based devices. Notably, this is achieved without any complicated microstructured patterns typical of SMSP fibers. *T*-Guides are thus a promising geometry for SMSP Si waveguides, which could have applications in high-quality on-chip lasers using Raman or Kerr nonlinearities in the mid-IR [21,22].

*2.3 Dispersion engineering*

The unique geometry of *T*-Guides also presents new opportunities for waveguide dispersion engineering, which is of importance to phase-sensitive nonlinear optical processes, such as four-wave mixing and supercontinuum generation. For *T*-Guides, dispersion engineering can be achieved by adjusting the overall size of the structure, as well as the relative width of the post and the thickness of the slab. Dispersion designs for silicon *T*-Guides are considered in Fig. 5, with sub-micron (i) and micron-sized (ii) geometries (see caption), showing the total group velocity dispersion (GVD) of each. Both designs show broad regions of small anomalous GVD, with varied zero-GVD wavelengths. In case (i), the anomalous-GVD region spans from wavelengths of 2.3 to 3.6 μm. In case (ii), it extends much further, from 2.9 to 7.8 μm. These results show promise for nonlinear integrated photonics with *T*-Guides, benefiting from their smooth dispersion profile and wideband SMSP operation.

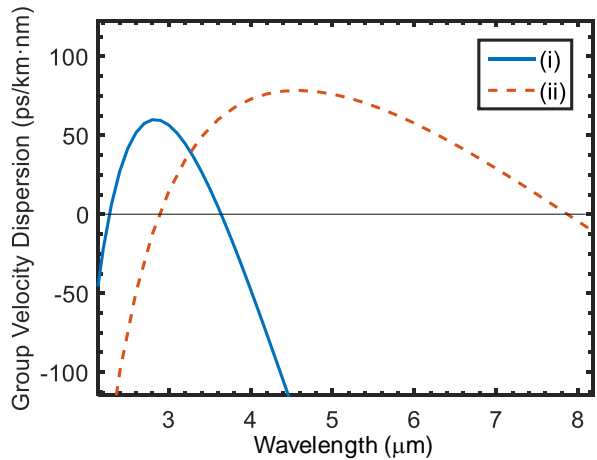

Fig. 5. Simulated total GVD of various silicon *T*-Guide designs: (i) $w = 0.85$ μm, $t = 0.64$ μm, and (ii) $w = 1.6$ μm, $t = 1.4$ μm.

## 3. Conclusion

In conclusion, anchored-membrane waveguides, or "*T*-Guides", have been numerically investigated as an integrated photonic platform supporting wideband single-mode and single-polarization SMSP operation. Air-clad silicon *T*-Guides were shown to exhibit SMSP windows of up to 2.75 octaves, a record span in simulations. Furthermore, dispersion engineering of *T*-Guides designs showed that promising dispersion profiles can be achieved with relatively flat and small anomalous GVD spanning from $\lambda = 2.9 – 7.8$ μm. Their superior SMSP bandwidth, smooth dispersion and simple geometry are ideally suited for nonlinear integrated photonics applications.

**Acknowledgments**

This work was supported by the National Science Foundation (NSF) CAREER (ECCS-1150672) and the Office of Naval Research (ONR) YIP (11296285). We thank Prof. Kathleen Richardson's research group for the use of their Firefly OPO, and Prof. Axel Schülzgen's group for the use of their 976 nm laser diode source.